# General Theory for Bilayer Stacking Ferroelectricity

Junyi Ji[1,2], Changsong Xu[1,2]*, and H. J. Xiang[1,2,3]*

[1]*Key Laboratory of Computational Physical Sciences (Ministry of Education), Institute of Computational Physical Sciences, and Department of Physics, Fudan University, Shanghai 200433, China*

[2]*Shanghai Qi Zhi Institute, Shanghai 200030, China*

[3]*Collaborative Innovation Center of Advanced Microstructures, Nanjing 210093, China*

Email: [csxu@fudan.edu.cn](csxu@fudan.edu.cn), [hxiang@fudan.edu.cn](hxiang@fudan.edu.cn)

**Abstract**

Two-dimensional (2D) ferroelectrics, which is rare in nature, enable high-density non-volatile memory with low energy consumption. Here, we propose a theory of bilayer stacking ferroelectricity (BSF), in which, two stacked layers of the same 2D material, with different rotation and translation, exhibits ferroelectricity. By performing systematic group theory analysis, we find out all the possible BSF in all the 80 layer groups (LGs) and discover the rules about the creation and annihilation of symmetries in the bilayer. Our general theory can not only explain all the previous findings (including sliding ferroelectricity), but also provide new perspective. Interestingly, the direction of the electric polarization of the bilayer could be totally different from that of the single layer. In particular, the bilayer could become ferroelectric after properly stacking two centrosymmetric non-polar monolayers. By means of first-principles simulations, we demonstrate that the ferroelectricity and thus multiferroicity can be introduced to the prototypical 2D ferromagnetic centrosymmetric material $CrI_3$ by stacking. Furthermore, we find that the out-of-plane electric polarization in bilayer $CrI_3$ is interlocked with the in-plane electric polarization, suggesting that the out-of-plane polarization can be manipulated in a deterministic way through the application of an in-plane electric field. The present BSF theory lays a solid foundation for designing a large number of bilayer ferroelectrics and thus colorful platforms for fundamental studies and applications.

**Main text**

Ferroelectrics are important functional materials and are widely used as sensors, actuators and information devices. The recent emergence of two-dimensional (2D) ferroelectrics, overcoming the size effects, leads to promising applications of high-density memory[1-3] and new optoelectronic devices[4-6]. So far, the number of experimentally determined 2D ferroelectrics is rather limited. The first example of 2D ferroelectrics is $CuInP_2S_6$, which is obtained from van der Waals layers and exhibits room-temperature ferroelectricity[7-9]. In 2016, SnTe films were synthesized to the limit thickness of one unit cell and exhibited robust in-plane ferroelectricity at nearly room temperature[10]. Later, another ferroelectric monolayer α-$In_2Se_3$ was prepared from its van der Waals bulk and was demonstrated to possess both out-of-plane and in-plane polarization. Moreover, an organic-inorganic hybrid ferroelectric $BA_2PbCl_4$[11] and a type-II multiferroic, $NiI_2$ [12] were recently reported. Despite such examples, the limited choices of 2D ferroelectrics strongly hinders fundamental studies and realistic applications.

Recently, sliding ferroelectricity was proposed to construct 2D ferroelectrics from non-ferroelectric monolayers. The sliding ferroelectricity refers to the phenomenon that ferroelectricity can arise from a bilayer, which is made of two monolayers, and the polarization can be switched by sliding one monolayer with respect to another. Such idea is theoretically proposed by Li and Wu in 2017, in the demo bilayers of BN and $MoS_2$[13]. The sliding ferroelectricity in BN was then experimentally observed by two groups in 2021. Yasuda *et al.* probed the polarization switching of parallel-stacked bilayer BN by measuring the resistance of an adjacently stacked graphene sheet [14]. Stern *et al.* realized the reversible switching of polarization by lateral sliding one of the monolayers using the biased tip of an atomic force microscopy [15]. Furthermore, the sliding ferroelectricity is also realized in bilayers of transition metal dichalcogenides, such as $WTe_2$ and $MoS_2$ [16-21]. However, current studies on sliding ferroelectricity are limited to specific systems and peculiar amount of sliding, leaving a large number of 2D systems unexplored. Hence, a general rule, that can govern the generation of ferroelectricity from stacking two monolayers, is highly desired.

In this Letter, by applying group theory over all the 80 layer groups (LGs), which cover all monolayers, we propose a more generalized concept of bilayer stacking ferroelectricity (BSF). Such theory can easily tell whether a bilayer exhibits electric polarization and what operations (e.g., rotation and shifting) are needed, with only the LG of monolayer as input. Particularly, the known sliding ferroelectrics BN, $MoS_2$ and $WTe_2$ can be easily understood within the BSF theory, which predicts additional types of electric polarizations in these systems. Moreover, ferroelectric bilayers are made from the centrosymmetric monolayer magnet $CrI_3$, thus making the bilayer a ferromagnetic-ferroelectric multiferroic system. More interestingly, a deterministic approach to flip out-of-plane polarization with in-plane electric field is proposed.

*Group theory analysis of BSF*. Any 2D material can be regarded as a single layer. The bilayer system can be seen as two single layers with one layer unmoved and the other layer rotated and translated by certain operations $\hat{O}$ [see Fig.1(a)]. For a single layer with a fixed symmetry group, different operations $\hat{O}$ may lead to different symmetry groups [see Fig.1(e) as an example] and thus abundant properties in the bilayer. We will derive the operations $\hat{O}$ that transform the symmetry group $G_S$ of a single layer to that $G_B$ of the bilayer. In this section, we first introduce the four polar types, which is closely related to the symmetries. Then we discuss the presence or absence of a certain symmetry in the single layer and the stacked bilayer. Finally, we apply the result of each symmetry to 80 LGs and figure out the group and polar type after stacking.

The symmetry of a 2D system is described by a LG, $G$. A symmetry operator $\hat{R}$ in $G$ gives a constraint on the polarization of the system. All these constraints determine the allowed polarization direction in the system. Therefore, the 80 LGs can be classified into four polar types [in-plane polarization (IP), out-of-plane polarization (OP), combined polarization with both in-plane and out-of-plane components (CP), non-polar (NP)]. There are 15 IP LGs, 13 OP LGs, 4 CP LGs and 48 NP LGs (see Table S1). After stacking, the polar type of a 2D system may be transformed to another polar type. In essence, this is because, after stacking, some symmetries are broken and new symmetries may emerge. Supposing the single layer 2D system $S$, which belongs to the LG $G_S = \{\hat{R}_S\}$, is parallel to the $xy$-plane. The second layer $S'$ comes from the operator $\hat{\tau}_z \hat{O}$ acting on $S$. Here, $\hat{\tau}_z$ is an arbitrary out-of-plane translation operator which does not affect the symmetry of the bilayer system $B = S + S'$, and $\hat{O} = \{O|\tau_o\}$ is the stacking operator that we mainly focus on, where $O$ is the rotational part and $\tau_o$ is the in-plane translational part. The bilayer, as shown in Fig. 1(a), with the LG $G_B = \{\hat{R}_B\}$, will satisfy a set of equations due to the presence of symmetry (eq. (2) in SM). If we solve this equation set directly, both $G_S, G_B$ can be one of the 80 LGs and we have to solve 6400 equation sets, which is very time consuming and provides not much physical insight. Here, we propose an easier and physically more appealing approach to solve them: (1) In the first step, we solve $\hat{O}$ in the equation set for a certain symmetry $R$ in any $G_B$. (2) In the second step, we apply the solution about $\hat{O}$ to single layer $S$ with specific symmetries. Since the number of $R$ is about 10 times smaller than the number of $G_B$, this method is almost 10 times faster than the direct solving approach.

Now let us discuss the first step of our approach. Since $\forall \hat{R}_S \in G_S, \hat{R}_S S = S$, we have $\hat{\tau}_z \hat{O} \hat{R}_S S = \hat{\tau}_z \hat{O} S = S'$. Therefore, $\hat{O} \hat{R}_S$ gives the same bilayer system as $\hat{O}$ for any $\hat{R}_S \in G_S$, thus for simplicity we use $\hat{O}$ to represent the solution set $\hat{O} G_S$ hereafter. To solve the equation set (eq. (2) in SM), we classify $\hat{R}$ ($\hat{R}_S, \hat{R}_B$) into two categories, $\hat{R}^{\pm}$, according to whether the rotational part, $R$, reverses the $z$-coordination (and the z component of polarization). And the rotational symmetry $R$ that reverses the $z$-coordination will exchange $S$ and $S'$. Those $R$ that can (or cannot) exchange $S$ and $S'$ are labeled with superscript " $-$ " (" $+$ "), including $\{R^-\} = \{I, m_z, c_{2\alpha}, s_{nz}\}$ ($\{R^+\} = \{E, m_\beta, c_{nz}\}$, $m_\beta$ is a mirror plane that is perpendicular to the $xy$-plane) (see SM for details). We define $G_{S0}(G_{B0})$ as the point group of the system $S(B)$. Considering all possible $R \in G_{B0}$, we divide the equations sets into four independent cases and solve them separately: (i) $R = R^- \in G_{S0}$, (ii) $R = R^- \notin G_{S0}$, (iii) $R = R^+ \in G_{S0}$, (iv) $R = R^+ \notin G_{S0}$. Our theoretical analysis results in several principles [see Fig. 1(a)~(d)]: (a) The $R^-$ symmetry in the bilayer $B$ must be one of the elements of $OG_{S0}$, i.e., $R_B^- \in$

$\{R^-\} \cap OG_{S0}$. The bilayer system $B$ has a given $R^-$ symmetry only if $O \in R^- G_{S0}$. In this case, $B$ has no out-of-plane polarization due to the presence of $R^-$. And one can choose $O \notin R^- G_{S0}$ to ensure the breaking of $R^-$ symmetry and presence of out-of-plane polarization in $B$. (b) Any pure translation $\hat{O} = \{E|\tau_o\}$ (including $\{E|\tau_o\}G_{S0}$) cannot break the inversion symmetry of the single layer $S$, suggesting that sliding ferroelectricity is impossible based on a centrosymmetric single layer. In contrast, a pure translation $\tau_o$ may break all other symmetries. (c) If the bilayer $B$ has the $R^+$ symmetry, such as $c_{nz}$ and $m_\beta$, then the original single layer $S$ must have the same $R^+$ symmetry. This indicates that any additional $R^+$ symmetry cannot be introduced by stacking. (d) For the single layer with the $c_{nz}$ symmetry that excludes the in-plane polarization, whether the $c_{nz}$ symmetry is kept in the bilayer $B$ is determined by the translational part $\tau_o$ but not the rotational part of $\hat{O}$.

These principles are general and suitable for all 2D bilayers, including the twisted bilayer systems that attracted numerous attentions recently. In 2021, Woods *et al.* observed triangular dipolar domains in marginally twisted hexagonal BN [22]. Recently, Weston *et al.* observed robust room temperature ferroelectricity in marginally twisted $MoS_2$ [23]. Here, the present BSF leads to several rules for realizing ferroelectricity in twisted bilayer system. We can prove that there is no inversion symmetry in the small angle twisted bilayer. To get a CP twisted bilayer from any single layer, one can rotate the second layer around the $z$ axis by a small angle with a general arbitrary translation. According to principles (a) and (b), all symmetries are broken and the bilayer belong to $C_1$. To get an IP twisted bilayer from any single layer, one can choose $\hat{O} \in \{c_{2\alpha}|\tau_\alpha\}G_{S0}$, where $c_{2\alpha}$ is a two-fold rotation along the axis $(\cos\alpha, \sin\alpha, 0)$ and $\tau_\alpha$ is a non-zero translation along the direction $(-\sin\alpha, \cos\alpha, 0)$. The bilayer will belong to LG 8~10 ($C_2$) and have an in-plane polarization (see SM for details). To get an OP twisted bilayer from a single layer, the single layer must have $c_{nz}$ symmetry according to principle (c). According to principle (a) and (b), one can rotated the second layer around the z axis by a small angle to breaking all the other symmetries. On the other hand, a certain translation will keep the $c_{nz}$ symmetry according to principle (d) [see Fig. 1(d)]. These three cases are illustrated in the SM using a single layer with the LG-80 as an example.

As we discussed above, twisting provides a general strategy to introduce ferroelectricity in the bilayer system. However, the large supercell size of the twisted bilayer indicates small electric polarizations. For simplicity and to enhance the electric polarization of the bilayer, we focus on the bilayer that has the same lateral unit cell size as the corresponding single layer. In this case, the 2D lattice remain unchanged under $\hat{O}$. Since $\hat{O}$ is equivalent to $\hat{O}\hat{R}_S$, we can infer that $O$ must be an element in the factor group $G_O = G_L/G_{S0}$, where $G_L$ is symmetry point group of the 2D Bravais lattice ($C_{2h}$ for LG 1~7, $D_{2h}$ for LG 8~48, $D_{4h}$ for LG 49~64, $D_{6h}$ for LG 65~80). We solve the equation sets for all 80 LGs and find out how each LG ($G_S$) of the single layer is transformed into the LG ($G_B$) of the bilayer under all possible stacking operations $\hat{O}$ (see SM for details). An example with the LG 66 is showed in Fig.1 (e). In Table I, we list how the four polar types transform by stacking for all LGs. Some key findings are summarized as follows: (1) The 2D system that has the same PG (i.e. the highest symmetry with a certain lattice) as its Bravais lattice cannot become ferroelectric through any stacking without expanding the lateral cell size. (2) For non-polar single layer with other non-polar PG, it is always possible to introduce in-plane polarization by stacking. (3) If the single layer has in-plane polarization, then the stacked bilayer cannot have nonzero out-of-plane polarization but vanishing in-plane polarization. (4) Out-of-plane ferroelectricity can be induced by stacking two centrosymmetric monolayers (e.g., the monolayers with the $C_{3h}$ symmetry).

Recently, the bilayer heterostructures obtained by stacking two different monolayers were found to display ferroelectricity. For example, $MoS_2/WS_2$ heterobilayer displays out-of-plane ferroelectricity [24] and $WSe_2$/Black phosphorus (BP) Van der Waals (vdw) heterointerface has an in-plane electronic polarization [25]. Here, we find that our previous theory on stacking two same monolayers can be easily extended to the heterostructure case. For bilayer heterostructure, the two components are represented by $S1$ and $S2$ with their LGs denoted as $G_{S1} = \{\hat{R}_{S1}\}$ and $G_{S2} = \{\hat{R}_{S2}\}$, respectively.

The bilayer heterostructure is represented by $B_h = S_1 + S_2' = S1 + \hat{\tau}_z \hat{O} S_2$ ($S1 \neq S2$). Since $S1$ and $S2'$ cannot be exchanged by any symmetries, there is no $\hat{R}^-$ symmetry in the heterostructure and only $\hat{R}^+$ symmetries are needed for consideration. Principles (c) and (d) are still valid here. Obviously, the trivial out-of-plane polarization may exist for all bilayer heterostructures. Therefore, we focus on the in-plane polarization of the heterostructure. If the two monolayers do not have a common $c_{nz}$ symmetry, the heterostructure will have an in-plane polarization. In contrast, if both monolayers have the same $c_{nz}$ symmetry, the $c_{nz}$ symmetry must be broken so as to induce the in-plane polarization. This can be achieved by interlayer sliding so long as the rotational axes of $c_{nz}$ in $S1$ and $S2$ don't coincide. In the WSe$_2$/BP vdw heterointerface, the two layers don't have the same $n$-fold rotation, which explains the presence of the in-plane polarization [25].

*BSF in BN and MoS$_2$.* In this part, we apply the above BSF theory to the known sliding ferroelectrics, such as BN, MoS$_2$ and WTe$_2$.

Monolayer BN and MoS$_2$ share the same symmetries, as their LG is $p-6m2$ (LG-78) and the corresponding PG is $D_{3h}$. According to Table I, it is possible for the bilayers to have all the four types of polarizations. The directions of induced polarizations and corresponding operations $\hat{O}$ (see details in the SM) are summarized in Table II. We now use the information from Table II to show how to generate different types of polarizations. For this LG, we find that one only needs to consider the case of sliding ferroelectricity (i.e., $O = E$). This can be reasoned as follows: (i) the $G_h$ for hexagonal systems is $D_{6h}$ and the factor group is $G_O = G_L/G_{S0} = D_{6h}/D_{3h} = \{E, I\}$; (ii) the principle (b) indicates that for any $\tau_o$ associated with $O = I$ will induce the inversion symmetry and forbid the polarization. Our analysis on sliding ferroelectricity for this LG leads to the following results: **(1)** Simply stacking two monolayers with no sliding and rotation leads to the bilayer being at high symmetry point $G$ [hereafter symmetry points denote the peculiar in-plane translations, see Fig. 2(b)], which belongs to $D_{3h}$ and has no polarization. **(2)** The high symmetry points $M$ and $N$ refer to the translation of starting from the simply stacked bilayer of case (1) and shifting the top layer by (1/3,2/3) and (2/3,1/3), respectively, in units of unit cell lattice vectors $(\boldsymbol{a}, \boldsymbol{b})$. The states at M and N points have $C_{3v}$ symmetry and exhibit polarization along [001] direction. Such case corresponds to the so-called AB and BA states of BN bilayer [15]. **(3)** The high symmetry points A, B and C points refer to shifting the top layer by (1/2,0), (0,1/2) and (1/2,1/2), respectively, and yields $C_{2v}$ symmetry for the bilayers. Such cases all exhibit in-plane polarization, but along [120], [210] and [1-10] respectively. Moreover, partially shifting from G to A, B and C leads to the high symmetry lines $GA, GB, GC$, which possess lower symmetry of $C_{2v}$. The polarization on such high symmetry lines is also in-plane, along the same directions as that at the ending high symmetry points. **(4)** For another group of high symmetry lines $GM, AN, GN, BM, MN$, the symmetry is further lowered to $C_s$ and the bilayer now have combined polarization with both in-plane and out-of-plane components. **(5)** Stackings at points other than the aforementioned high symmetry points and lines results in $C_1$ symmetry and the polarization could point toward any directions. According to items (1)-(5) from our BSF theory, the bilayer BN and MoS$_2$ not only can have pure out-of-plane polarization that was revealed by the theory of sliding ferroelectricity, but also have many more possibilities to exhibit pure in-plane and combined polarizations. Note that the analyses here apply to any systems that share $p-6m2$ layer group with BN and MoS$_2$, e.g., monolayer VGe$_2$N$_4$ [26]. Our theory is also applicable to bilayer WTe$_2$, as detailed in SM.

**BSF in CrI$_3$**. We now apply the BSF theory and DFT calculations to CrI$_3$, which is a typical 2D ferromagnets[27-29]. Monolayer CrI$_3$ belongs to layer group of $p-31m$ (LG-71) and point group of $D_{3d}$ with inversion symmetry. According to Table I, bilayer CrI$_3$ is possible to form the phases that exhibit IP or CP, but not pure OP. According to Table II, such polar states happen in the case that the top layer flips ($O = m_z$), while the simply stacked bilayer exhibits no polarization, due to the maintained inversion symmetry. Table II further indicates that the phases with IP emerge at A, B and C points, as well as along GM, AN, GN, BM and MN lines, while the states with CP arise along GA, GB and GC lines. DFT calculations are further performed to identify the energy minima of different stackings and the values of polarizations. As shown in Fig. 2(d), the simply stacked bilayer [G point, see Fig. 2(b)], which exhibits no polarization and corresponds to the energy maximum; While shifting the top layer by ±1/3 along the directions of along GA, GB and GC lines leads to the six-fold degenerate global minima with combined polarizations. The energy gain from maximum to minima yields 47.98 meV/f.u., indicating good stability of the CP states at minima. The polarization at, e.g., (1/3,0) point yields 0.1482 μC/cm$^2$, with an in-plane component of 0.1475 μC/cm$^2$ and an out-of-plane component of 0.0146 μC/cm$^2$. Specifically, the polarizations at GM, AN and A point toward [100], while those at GN, BM and B point toward [010]. Notably, Fig. 2(d) further reveals that it doesn't have to go across the energy maximum of G point to switch among the degenerate CP ground states. Two types of paths are found to be interesting: (i) directly from (0,-1/3) to (-1/3,-1/3), which not only flips the OP component but also rotate the IP component by 120 degrees, and which only needs to overcome a small energy barrier of 6.36 meV/f.u.; and (ii) directly from (1/3,0) to (2/3,0), which only flips the OP component but leaves the IP component unchanged, and which yields a moderate energy barrier of 13.88 meV/f.u.. Such results indicate that the BSF theory is well consistent with DFT calculations, and that one has to rely on DFT to identify the polarization values, locate energy minima and find the suitable transition path.

Moreover, the BSF theory and DFT results also indicate a controllable approach to switch out-of-plane polarization with in-plane electric field. Let us focus on path (i) that goes directly from (0,-1/3) to (-1/3,-1/3). Along such path, the direction of IP goes from [0-10] to [110], which indicates that an electric field along [120] (perpendicular to the path, but parallel to the direction of the in-plane polarization difference) is possible to complete such path and that the OP flipping can also be accomplished. Such approach is indeed feasible as that, (i) with an electric field along [120], the shifting from (0,-1/3) to (-1/3,-1/3) corresponds to the lowest energy barrier and it is the best path to lower the interaction energy with the form of $-\boldsymbol{E} \cdot \boldsymbol{P}$; and (ii) that the IP component is ten times larger than OP component. Such switching mechanism, which utilizes the locking between in-plane and out-of-plane polarization, is similar with the dipole locking enabled switching in In$_2$Se$_3$ [30]. On the other hand, such approach with in-plane field does not work for path (ii), as that it does not change the IP direction from (1/3,0) to (2/3,0). Note that applying out-of-plane electric field is found to be an uncontrollable way to tune polarization, since each CP ground state corresponds to three shifting directions, with two degenerate path (i) and one path (ii). Similarly, it is found impossible to switch deterministically the polarization of BN using an external electric field, as the so-called AB and BA stackings exhibit only out-of-plane polarization and correspond to three-fold degenerate switching paths (See Fig. S2).

The bilayer $CrI_3$, at its ground states, is further found to be a rare ferromagnetic and ferroelectric multiferroic system. The energy difference between interlayer FM state and AFM state is calculated for each shifting. As revealed by Fig. 2c, the magnetic ground states yield FM at the six-fold degenerate CP states, while the NP states at G, M and N points correspond to AFM states. Note that a similarly stacked $CrBr_3$ bilayer (H-type stacked) with FM state has been prepared [31], indicating a good possibility for the realization of multiferroicity in the presently proposed $CrI_3$ bilayer. Such FM-vs-AFM states at different stacking conditions are consistent with previous studies [32-34], and it thus enables multi-state storage, together with the presently demonstrated BSF.

***Summary***. To conclude, the BSF theory is proposed to guide the construction of polar bilayers from monolayers. Typical sliding ferroelectrics of BN, $MoS_2$, $WTe_2$ are re-examined and are found to be specific cases of the present BSF theory. Combining BSF theory and DFT calculations, the bilayer $CrI_3$, breaking the inversion symmetry of its monolayers, is demonstrated to exhibit both in-plane and out-of-plane polarizations, yielding a ferromagnetic-ferroelectric multiferroic system. An approach is also proposed to flip out-of-plane polarization with in-plane electric field in bilayer $CrI_3$. We thus hope that many novel ferroelectric systems can be observed with the guidance of our BSF theory.

**Acknowledgement**. This work is supported by NSFC (grants No. 811825403, 11991061, 12188101, 12174060, and 12274082) and the Guangdong Major Project of the Basic and Applied Basic Research (Future functional materials under extreme

conditions--2021B0301030005). C.X. also acknowledge supports from the open project of Guangdong provincial key laboratory of magnetoelectric physics and devices (No. 2020B1212060030).

Table. I. Categories of the 80 LGs, according to the polar types of the bilayers. $P_S$ and $P_B$ denote the monolayer and bilayer polar status, respectively. The labels "√" and "×" represent feasible and infeasible, respectively.

| $P_S$ | | IP | OP | | CP |
|---|---|---|---|---|---|
| $G_S$: LG (PG) | | 4,5($C_s$), 8~10($C_2$) 27~36($C_{2v}$) | 3($C_2$), 23~26($C_{2v}$), 55,56($C_{4v}$), 77($C_{6v}$) | 49($C_4$), 65($C_3$) 69,70($C_{3v}$), 73($C_6$) | 1($C_1$) 11~13($C_s$) |
| $P_B$ | IP | √ | × | √ | √ |
| | OP | × | √ | √ | × |
| | CP | √ | √ | √ | √ |
| | NP | √ | √ | √ | √ |

| $P_S$ | | | NP | | |
|---|---|---|---|---|---|
| $G_S$: LG (PG) | | 6,7($C_{2h}$), 37~48($D_{2h}$) 61~64($D_{4h}$), 80($D_{6h}$) | 2($C_i$), 14~18($C_{2h}$), 19,21($D_2$) 51,52($C_{4h}$), 53,54($D_4$), 57,58($D_{2d}$) 71($D_{3d}$), 75($C_{6h}$), 76($D_6$), 79($D_{3h}$) | | 22($D_2$), 50($S_4$), 59,60($D_{2d}$) 66($C_{3i}$), 67,68($D_3$), 72($D_{3d}$), 74($C_{3h}$), 78($D_{3h}$) |
| $P_B$ | IP | × | √ | | √ |
| | OP | × | × | | √ |
| | CP | × | √ | | √ |
| | NP | √ | √ | | √ |

Table II. The directions of polarizations at different stackings, for BN, MoS₂ and CrI₃. Here we assume that the bottom layer is static and top layer is governed by the shifting defined by high symmetry points and lines. The rule that stackings does not change the size of unit cell applies here. The coordinates of high symmetry points are in unit of lattice vectors (**a**,**b**). Point groups (PG) are also indicated in this table.

| High symmetry points | Coordinates | $BN, MoS_2$ $O = E$ | | $CrI_3$ $O = m_z$ | |
|---|---|---|---|---|---|
| | | PG | $\vec{e}_P$ | PG | $\vec{e}_P$ |
| G | (0,0) | $D_{3h}$ | × | $D_{3h}$ | |
| M | (1/3,2/3) | $C_{3v}$ | 001 | $D_3$ | × |
| N | (2/3,1/3) | | | | |
| A | (1/2,0) | $C_{2v}$ | 120 | $C_{2v}$ | 100 |
| B | (0,1/2) | | 210 | | 010 |
| C | (1/2,1/2) | | 1-10 | | 110 |
| High symmetry lines | | | | | |
| GA | | $C_2$ | 120 | $C_S$ | 100, 001 |
| GB | | | 210 | | 010, 001 |
| GC | | | 1-10 | | 110, 001 |
| GM, AN | | $C_S$ | 120, 001 | $C_2$ | 100 |
| GN, BM | | | 210, 001 | | 010 |
| MN | | | 1-10, 001 | | 110 |
| General points | | $C_1$ | all | $C_1$ | all |

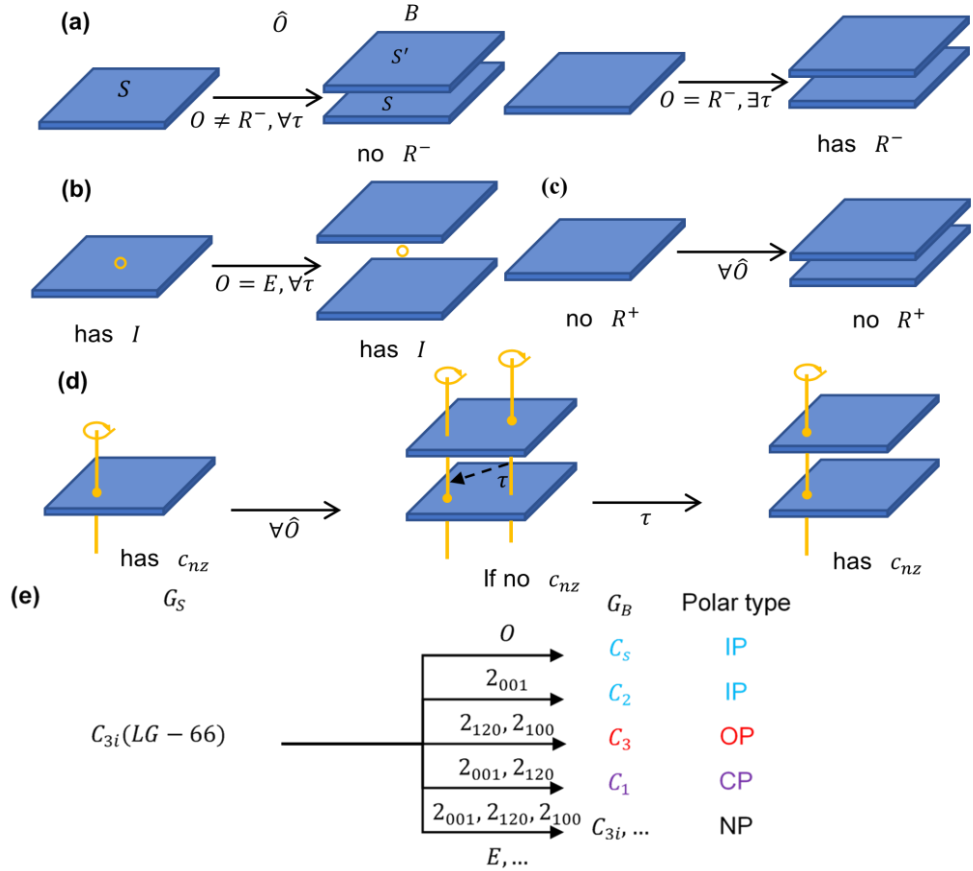

Figure 1. Schematization of constructing bilayers from monolayers and illustration of the four principles in the text. (a) the $\hat{R}^-$ symmetry $(I, m_z, c_{2\alpha}, s_{nz})$ can exist in $B$ only if $O \in R^- G_{S0}$. If $O \notin R^- G_{S0}$, there won't be $R^-$ in $B$. (b) One case of principle (b): if the single layer is centrosymmetric and the bilayer is stacked with pure translation, the bilayer will also be centrosymmetric. (c) principle (c): if the single layer does not have $R^+(E, m_\beta, c_{nz})$, the bilayer also does not have this $R^+$. (d) interpretation of principle (d): the stacking progress will always keep the n-fold rotational axis (colored in orange) parallel to the z-axis, thus a translation that coincide one n-fold rotational axis in both layers will preserve the n-fold rotation symmetry. (e) example of the polar type transformation after bilayer stacking: the single layer belongs to $G_S$ and the second layer is rotated by different $\hat{O}$. With different rotational part $O \in G_O$, the bilayers can belong to different $G_B$ and have four polar types: in-plane-polar (IP), out-of-plane-polar (OP), combined-polar (CP), and non-polar (NP) groups. The translational part is omitted here.

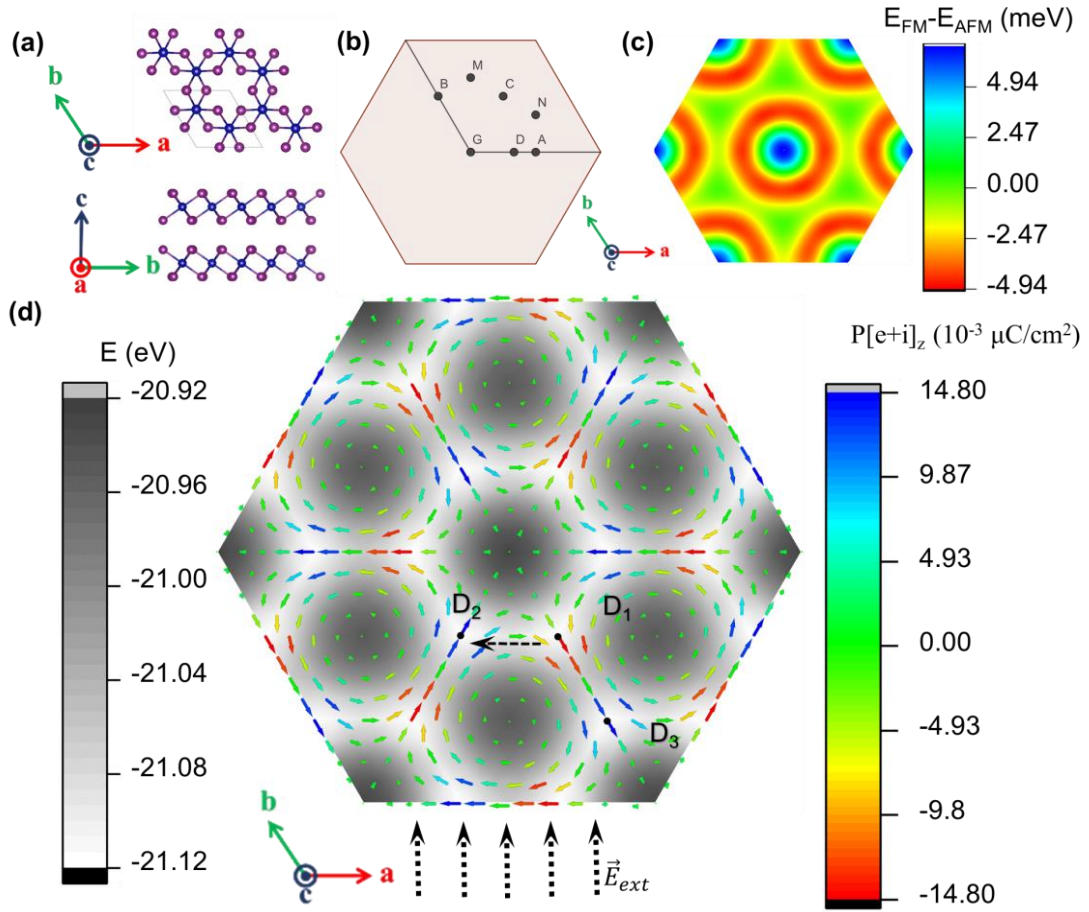

Figure 2. (a) Top and side view of bilayer CrI3 rotated by $\hat{O} = \{m_z|0\}$. Purple and blue spheres denote I and Cr atoms, respectively. (b) The six high symmetry points in the "unit cell" of $\tau_0$ in the hexagonal crystal system and the minima point D in CrI3. The parallelogram with black edges is the "unit cell" of $\tau_0$. (c) The magnetic states distribution of the magnetic ground states in the "unit cell". (d) The polarization and energy distribution of the magnetic ground states in the "unit cell". The global ground states around the original point are at $D_{1,2,3...}$: $\pm(1/3,0), \pm(0,1/3), \pm(1/3,1/3)$. The in-plane and the out-of-plane polarization components are represented by the solid arrows and the color of the solid arrows, representatively. The dotted arrows and dashed arrows show the external electric field $\vec{E}_{ext}$, which is in-plane and very small, and the switching pathway under the external electric field.